c9906.tex

\newcommand{\be}{\begin{equation}}
\newcommand{\ee}{\end{equation}}
\newcommand{\bea}{\begin{eqnarray}}
\newcommand{\eea}{\end{eqnarray}}
\newcommand{\beann}{\begin{eqnarray*}}
\newcommand{\eeann}{\end{eqnarray*}}
\newcommand{\beasn}{\begin{sneqnarray}}
\newcommand{\eeasn}{\end{sneqnarray}}
\newcommand{\ba}{\begin{array}}
\newcommand{\ea}{\end{array}}
\newcommand{\nn}{\nonumber}
\newcommand{\Appendix}[1]%
    {\renewcommand{\thesection}{Appendix~\Alph{section}:}%
     \section{#1}%
     \renewcommand{\thesection}{\Alph{section}} }

\catcode`@=11
\def\secteqno{\@addtoreset{equation}{section}%
\def\theequation{\thesection.\arabic{equation}}}
\def\endsecteqno{\def\theequation{\@ifundefined{chapter}%
{\arabic{equation}}{\thechapter.\arabic{equation}}}}
\newcounter{subequation}
\def\thesubequation{\alph{subequation}}
\def\sneqnarray{\stepcounter{equation}\let\@currentlabel=\theequation
\setcounter{subequation}{1}
\def\@eqnnum{{\rm (\theequation\thesubequation)}}
\global\@eqcnt\z@\tabskip\@centering\let\\=\@eqncr\let\@@eqncr=\@@sneqncr
$$\halign to \displaywidth\bgroup\@eqnsel\hskip\@centering
 $\displaystyle\tabskip\z@{##}$&\global\@eqcnt\@ne
 \hskip 2\arraycolsep \hfil${##}$\hfil
 &\global\@eqcnt\tw@ \hskip 2\arraycolsep
$\displaystyle\tabskip\z@{##}$\hfil
  \tabskip\@centering&\llap{##}\tabskip\z@\cr}
\def\endsneqnarray{\@@sneqncr\egroup $$\global\@ignoretrue}
\def\@@sneqncr{\let\@tempa\relax
   \ifcase\@eqcnt \def\@tempa{& & &}\or \def\@tempa{& &}
   \else \def\@tempa{&}\fi
     \@tempa \if@eqnsw\@eqnnum\stepcounter{subequation}\fi
     \global\@eqnswtrue\global\@eqcnt\z@\cr}
\def\nobiblabels{\def\@lbibitem[##1]##2{\@bibitem{##2}}}
\catcode`@=12

   \def\g{\gamma} \def\G{\Gamma}

   \def\m{\mu} 

\def\s{\sigma} \def\t{\tau}  
    
\def\o{\omega} \def\O{\Omega} 

\newcommand{\PRL}[3]{{\it Phys. Rev. Lett.} {\bf #1} (19#2) {#3}}

\newcommand{\PR}[3]{{\it Phys. Rev.} {\bf #1} (19#2) {#3}}

\newcommand{\JPCM}[3]{{\it J. Phys.: Cond. Matter} {\bf #1} (19#2) {#3}}
 
\def\sv{\mbox{\boldmath $\sigma$}}
\def\Mt{{\bf M}}

\def\pa{\partial} \def\da{\dagger} 

\documentclass[12pt]{article}
\usepackage{epsfig}

\secteqno
\catcode`@=11
\long\def\@makecaption#1#2{
   \vskip 10pt
   \setbox\@tempboxa\hbox{{\small\bf #1.} \ {\small #2}}
   \ifdim \wd\@tempboxa >\hsize       
   {\small\bf #1.} \ {\small #2}\par  
   \else                              
        \hbox to\hsize{\hfil\box\@tempboxa\hfil}
   \fi}
\catcode`@=12

\begin{document}


\title{{\bf Continuum Double Exchange Model}}
\author{{\Large {\sl Jos\'e Mar\'{\i}a Rom\'an}
               \ and \  {\sl Joan Soto}}\\
        \small{\it{Departament d'Estructura i Constituents
               de la Mat\`eria}}\\
        \small{\it{and}}\\
        \small{\it{Institut de F\'\i sica d'Altes Energies}}\\
        \small{\it{Universitat de Barcelona}}\\
        \small{\it{Diagonal, 647}}\\
        \small{\it{E-08028 Barcelona, Catalonia, Spain.}}\\
        {\it e-mails:} \small{roman@ecm.ub.es, soto@ecm.ub.es} }
\date{\today}

\maketitle

\thispagestyle{empty}

\begin{abstract}
We present a continuum model for doped manganites which consist of two species 
of quantum spin $1/2$ fermions interacting with classical spin fields. The 
phase structure at zero temperature turns out to be considerably rich: 
antiferromagnetic insulator, antiferromagnetic two band conducting, canted 
two band conducting, canted one band conducting and ferromagnetic one band 
conducting phases are identified, all of them being stable against phase 
separation. There are also regions in the phase diagram where phase 
separation occurs.  
\end{abstract}

\bigskip

PACS: 75.30.Et, 75.25.+z, 11.30.Qc, 75.10.-b 

\vfill
\vbox{
\hfill{cond-mat/9810389
}\null\par
\hfill{UB-ECM-PF 98/18
}\null\par}

\newpage


\section{Introduction}
\indent

Doped manganites $La_{1-x}A_{x}MnO_3$ ($A$ divalent) \cite{Jonker} 
are receiving quite a lot of both theoretical 
\cite{Paco,Paco2,Kagan,Zou,Millis,Maezono,Golosov,Dagotto} and experimental 
\cite{ref7&8Dagotto} attention lately. These materials show an 
interesting interplay between magnetism and conductivity with intrincated 
phase diagrams which are still controversial.

\medskip

In a cubic lattice  the $3d$ orbitals of $Mn$ split into a $t_{2g}$ 
triplet and an upper $e_{g}$ doublet. Due to the electronic repulsion and 
the Fermi statistics (Hund's
rule) the three $t_{2g}$ levels are always single occupied forming a core 
$S=3/2$ spin. The $e_{g}$ orbitals may be further splitted by a static 
Jahn-Teller distortion at small doping \cite{Zener}. 

\medskip

The above features are encoded in the so called double exchange models of
different degrees of complexity. The simpler ones assume a strong 
Jahn-Teller distortion so that only the lower $e_{g}$ level is consider. 
Hence there is a single fermion field in each site, with a spin independent 
hopping term and a local interaction with the core spin \cite{Kagan,Zou}. 
Core spins also interact among themselves with the usual Heisenberg 
term. Under certain assumptions \cite{Anderson} the interaction with the 
core spin can be traded for an angle dependent hopping term 
\cite{Paco, DeGennes}. The next level of complexity consist of taking 
into account the two $e_{g}$ levels \cite{Millis,Maezono}, and only very 
recently, the Jahn-Teller distortion has been incorporated dynamically 
by some \mbox{authors \cite{Dagotto}.}

\medskip

It is the aim of this work to present a simple continuum model for doped 
manganites which also encodes the basic features above and, moreover, is 
exactly solvable for classical core spins. It produces a rich phase diagram
which is in qualitative agreement with recent results and it shows, in 
addition, that stable canted phases exist. The main advantage with respect to 
previous approaches is that all the parameters of the material (lattice 
spacing, band curvature, Hund coupling, Heisenberg coupling and doping) 
combine into only two constants. This allows to present a two dimensional
phase diagram which holds for a large amount of materials.



\section{The Model} 
\indent

Cooperative phenomena are amenable of a field theoretical description. When the
phenomena do not depend on the details of the microscopic system but only on 
its long wave length behaviour a continuum field theory description is 
appropriated. The field theoretical continuum model must contain the relevant 
degrees of freedom at long wavelengths, which depend on the particular systems 
and phenomena that are to be studied. In our case, these are doped manganites 
and their phase diagram at zero temperature. These systems are 
known to undergo a number of phase transitions when the doping is increased. 
They are insulating antiferromagnets ($AFI$) at zero doping and become 
conducting ferromagnets ($FC$) at large enough doping. What happens between 
these two regimes is still controversial, though most of authors agree
that the phase diagram is very rich and non-trivial. Early works on the
subject suggested that an interesting intermediate conducting canted phase 
exists \cite{DeGennes}, but recent experimental \cite{ref7&8Dagotto} and 
theoretical \cite{Paco, Kagan} results indicate that the canted phase appears 
to be unstable against phase separation. 

\medskip

Theoretical work on the subject is 
based on variations of the double exchange models. The phase structure of the 
system is obtained from these models using certain simplifying assumptions 
(slave boson formalism \cite{Paco}, trial wave functions \cite{Kagan},\ldots) 
or extensive numerical simulations \cite{Dagotto}, the scope of which is 
difficult to evaluate. We present below a continuum field theoretical model 
which, as we shall argue, contains the relevant long wavelength degrees of 
freedom of the system. Then our main assumption is going to be that the rich 
phase diagram of manganites can be understood from long wavelength physics 
only. As the model is exactly solvable, there are no further uncertainties 
due to uncontrolled approximations.

\medskip

Since we wish our model to include the well established $AFI$ and $FC$ phases, 
we need at least an $AF$ order parameter field, a $F$ order parameter field, 
and a $I-C$ order parameter field. For the $AF$ and $F$ order parameter fields 
we shall 
use ${\bf M}_{1}(x)$ and ${\bf M}_{2}(x)$ the local magnetisations in the even 
and odd sublattices respectively. Both in the $AF$ and $F$ phases these local 
magnetisations are smoothly varying fields. In the $AF$ phase 
${\bf M}_{1}(x){\bf M}_{2}(x)\sim -1$ whereas in the $F$ phase 
${\bf M}_{1}(x) {\bf M}_{2}(x)\sim 1$. For the $I-C$ order parameter one
could think of introducing a single slowly varying spin $1/2$ fermion field 
together with a chemical potential which regulates the doping. When the 
chemical potential is below the energy gap of the lowest spin state we have 
an $I$ phase, when it overtakes this energy gap we have a one band $C$ phase, 
and when it overtakes the energy gap of the highest spin state we have a two 
band $C$ phase. However, a spin $1/2$ field naturally couples to the local 
magnetisation, which changes abruptly from the even to the odd sublattice in 
the $AF$ phase. Hence in this phase a single spin $1/2$ field cannot be slowly 
varying over the system. We need at least two slowly varying spin $1/2$ 
fermionic fields,  $\psi_1(x)$ which couples to the magnetisation in the even
sublattice ${\bf M}_{1}(x)$ and $\psi_2(x)$ which couples to the magnetisation 
in the odd sublattice ${\bf M}_{2}(x)$. Since the conductivity is due 
to fermions moving from one sublattice to the other one a (spin independent) 
hopping term is introduced. The allowed values of the chemical potential will 
be limited by the physical condition that no conduction must exist when the 
hopping parameter vanishes.  

\medskip

The model must be $SU(2)$ spin invariant since the magnetic interactions 
emerge from the usual superexchange mechanism together with the Hund's rule. 
The space-time symmetries of the underlying crystal must also be implemented 
and will be the only remain of the microscopic lattice structure. For 
simplicity we shall take a cubic lattice and comment later on the slight 
modifications that occur for other crystals.

\medskip

The lagrangian of the model reads
\bea
{\cal L}(x) & = & \psi^{\da}_1 (x) \left[ (1+i\epsilon)i\pa_0 + 
    {\pa^2_i \over 2m} + \m + J_H {\sv \over 2} \Mt_1 (x) \right] \psi_1 (x)  
\nn \\
& & \mbox{} + \psi^{\da}_2 (x) \left[ (1+i\epsilon)i\pa_0 + 
    {\pa^2_i \over 2m} + \m + J_H {\sv \over 2} \Mt_2 (x) \right] \psi_2 (x)  
                                        \label{lagrangian} \\
& & \mbox{} + t \left( \psi^{\da}_1 (x) \psi_2 (x) +  
           \psi^{\da}_2 (x) \psi_2 (x) \right) 
- J_{AF} \Mt_1 (x) \Mt_2 (x).  \nn
\eea

\medskip

The size of the parameters in the model are estimated by comparing them with 
the na{\"\i}ve continuum limit of lattice double exchange models. For a cubic 
lattice we have $2m\sim 1/a^2t^{l}$, $t\sim zt^{l}$, $J_{H}\sim J_{H}^{l}$ 
and $J_{AF}\sim  zJ_{AF}^{l}/a^3>0$, where $a$ is the lattice spacing, $z=6$
is the coordination number and the 
superscript $l$ means the analogous lattice quantity. 
The fields $\psi_{i}(x)$ may describe either electrons or holes. Since
the conduction in actual doped manganites is due to holes, one should better
figure out $\psi_{i}(x)$ as hole annihilating fields. Recall that for holes
$J_{H}$ is negative whereas it is positive for electrons. This sign however is
going to be irrelevant as far as the phase diagram is concerned.
  
\medskip

The lagrangian above is invariant under the following  transformations:
\begin{enumerate}
\item[(i)] Global $SU(2)$ spin transformations,
\be
\ba{rcl}
\psi_i (x) & \longrightarrow & g \psi_i (x)  \\
M^a_i (x) & \longrightarrow & R^a_b M^b_i (x),
\ea
\qquad \quad (i = 1,2)
\ee
\item[(ii)] Primitive translations,
\be
\ba{rlccrlc}
\psi_1 (x) & \longrightarrow & \psi_2 (x) & \quad & 
                               \psi_2 (x) & \longrightarrow \psi_1 (x) \\
\Mt_1 (x) & \longrightarrow & \Mt_2 (x) & \quad & 
                               \Mt_2 (x) & \longrightarrow \Mt_1 (x),
\ea
\ee
\item[(iii)] Point group transformations, given by the group $m\bar 3m$
\be
\ba{rcl}
\psi_i (x) & \longrightarrow & g_{\xi} \psi_i (\xi^{-1}x)  \\
M^a_i (x) & \longrightarrow & R^a_b(\xi) M^b_i (\xi^{-1}x)
\ea
\qquad \quad (i = 1,2),
\ee
when the point group transformation $\xi$ maps points in the same sublattice, 
and
\be
\ba{rlccrlc}
\psi_1 (x) & \longrightarrow & g_{\xi}\psi_2 (\xi^{-1}x) & \quad & 
                    \psi_2 (x) & \longrightarrow g_{\xi}\psi_1 (\xi^{-1}x) \\
M^a_1 (x) & \longrightarrow & R^a_b(\xi)M^b_2 (\xi^{-1}x) & \quad & 
                    M^a_2 (x) & \longrightarrow R^a_b(\xi)M^b_1 (\xi^{-1}x),
\label{point}
\ea
\ee
when the transformation $\xi$ maps points of different sublattices. Anyway, 
the rotations $g_{\xi}$ and  $R^a_b(\xi)$ can be absorbed by a $SU(2)$ 
transformation and the change of sublattice in (\ref{point})
by a primitive translation. Hence, in practice, we only have to care about the
transformation of the coordinates.
\item[(iv)] Time reversal,
\be
\ba{rcl}
\psi_i (x) & \longrightarrow & C \psi^*_i (Tx)  \\
\Mt_i (x) & \longrightarrow & - \Mt_i (Tx)
\ea
\qquad C = e^{-i\pi \s^2/2} = -i\s^2 \quad,\quad (i=1,2),
\ee
where $Tx=(-t, {\bf x})$.
\end{enumerate}


\section{Effective Potential}
\label{potential}
\indent

In order to find out how the ground state of the system changes as a function 
of the chemical potential, we shall calculate the effective potential and 
minimise it with respect to the order parameters ${\bf M}_1$ and 
${\bf M}_2$. We shall assume that the ground state configuration
corresponds to constant magnetisations both in the odd and even sublattices. 
Hence, the effective potential is to be minimised with respect to  the angle 
$\theta $ between ${\bf M}_1$ and ${\bf M}_2$ only. We use 
$y=\cos(\theta/2)$. When $y=0$, $0<y<1$ and $y=1$ we have an 
antiferromagnetic, canted and ferromagnetic phase respectively.

\medskip

The effective potential is obtained by integrating out the fermion fields in
the path integral, and it is formally given by
\be
V_{{\it eff}}=J_{AF}{\bf M}_1 {\bf M}_2  + i tr \log{\hat O}/VT,
\label{veff}
\ee 
where
\be
{\hat O} = 
\left( \ba{cc}
(1+i\epsilon)i\pa_0 + {\pa^2_i/2m} + \m + {J_H \over 2} \sv \Mt_1 
& t \\
t
& (1+i\epsilon)i\pa_0 + {\pa^2_i/2m} + \m + {J_H \over 2} \sv \Mt_2 
\ea \right),
\label{ohat}
\ee
and the trace is both on spin indices and space-time coordinates. $VT$ 
is the volume of the space-time. 

\medskip

If $\hat O$ has eigenvalues $\lambda_{n}$
\be
tr\log \hat O = \sum_{n} \log \lambda_{n}.
\label{sum}
\ee
We have then to diagonalise the operator $\hat O$. Since it contains only 
constant fields the diagonalisation with respect to the space-time is 
trivially attained by plane waves. The diagonalisation with respect to the 
spin indices is a simple linear algebra problem. We obtain
\bea
\lambda_{n} = O_{i}(q) & = & (1+i\epsilon )\;\o -{{\bf k}^2 \over 2m} -\O_{i}
\label{eigen}  \\ 
& & \nn \\
\O_{i} & = & \pm {\vert J_H\vert M \over 2} \;
     \sqrt{ 1 + \g^2 \pm 2\g \cos {\theta \over 2}} - \m
\quad, \quad \g \equiv {2t \over \vert J_H\vert M}. \label{el}
\eea
$q=(\o , {\bf k})$ and $M=\vert \Mt_1 \vert=\vert \Mt_2\vert=3/2$. 
The restriction for the values of the chemical potential in the model implies
that at most the two lower eigenvalues in (\ref{el}) may contribute.
This motivates the following reparametrisation of the chemical potential:
\be
\m = - {\vert J_H\vert M \over 2} \; \sqrt{ 1 + \g^2 - 2\g y_0}
\qquad\quad
(-1<y_0<y_0^{max}=\g/2),
\label{y_0def}
\ee
which eases comparison with the energy levels in (\ref{el})
($y = \cos(\theta/2)$).
 In order to simplify the analysis we assume $\g$ small 
and keep only linear terms in $\g$ in the relevant eigenvalues above. Namely,
\be
\O_i = - {\vert J_H\vert M \over 2} \;\g \;(y_0 \pm y).
\label{eigena}
\ee
This is justified for $t\ll J_{H}$, as it turns out to be the case for the 
actual materials \cite{Ramirez}. Anyway, this simplification can be lifted 
with the only drawback that the few analytic expressions below must also be 
substituted by numerical analysis. 

\medskip

In order to calculate the sum (\ref{sum}) we have used $\zeta$-function 
techniques \cite{Eli}, which are explained in the appendix. We obtain
the effective potential (for $\m <0$)
\be
V_{{\it eff}} =
V_0 \left[ (2y^2-1) - A \left( (y_o+y)^{5/2}\theta(y_0+y) +
                            (y_0-y)^{5/2}\theta(y_o-y) \right) \right],
\label{ep}
\ee
where we have defined
\be
V_0 = J_{AF}M^2 \qquad, \qquad 
A = {(2m)^{3/2}\; t^{5/2} \over 15 \pi^2 J_{AF}M^2}
= {z^{3/2} \over 15\pi^2}\;{t \over (J_{AF}a^3M^2)}.
\label{defs}
\ee


\section{Phase Structure}
\indent

The possible phases of the model are obtained by minimising (\ref{ep}) with 
respect to $y$ for the different values of the parameters $A$ and $y_0$. 
The number of conducting bands is given by the number of $\theta$-functions 
in (\ref{ep}) which contribute to the effective potential at the minimum.

\medskip

In order to gain some qualitative understanding and to make the minimisation 
procedure systematic we shall first separate the cases $y_0<0$ and $y_0>0$. 
For each case we shall work out the stability conditions for $AF$ ($y=0$),
canted ($0<y<1$) and $F$ ($x=1$) phases. After that we shall compare the 
energy of the stable phases and obtain the curves which separate them. 

\medskip

The stability conditions are given for the different phases by
\bea
AF: & & V_{{\it eff}}^{\prime}(0)>0 \quad or \quad  
        V_{{\it eff}}^{\prime}(0)=0 \;\; V_{eff}^{\prime\prime}(0)>0 \nn \\
C: & &  V_{{\it eff}}^{\prime}(y_c)=0 \;\; V_{eff}^{\prime\prime}(y_c)>0 \\
F: & &  V_{{\it eff}}^{\prime}(1)<0. \nn
\eea 

\medskip

Let us then consider first the case $y_0<0$. Clearly for $y_0<-1$ the 
unique existing phase is the $AFI$ phase. In the case $-1<y_0<0$ only the 
lowest
of the four spin eigenvalues may contribute to the effective potential.
The stability conditions yield the following stable phases:
\bea
AFI: & & y=0  \nn \\
FC:  & & y=1 \qquad A(1+y_0)^{3/2} > 8/5.  
\eea

\medskip

The canted phase is not stable as it can be seen from the condition 
$V_{{\it eff}}^{\prime}(y_c)=0$,
\be
y_c = {5 \over 8} A (y_0 + y_c)^{3/2},
\ee
which has at most one solution $y_c \in [-y_0,1]$. Since $V_{\it eff}$ is
continuous, and increasing at $y=0$ this solution must be a maximum when it
exists.

\medskip

The curve $V_{{\it eff}}(0)=V_{{\it eff}}(1)$ in the plain $(y_0,A)$, which 
separates the $AF$ and $F$ phases,  reads
\be
A(1+y_0)^{5/2} = 2 \qquad (-1<y_0<0).
\ee
Above this curve the $F$ phase is favoured against the $AF$ phase and 
viceversa.

\medskip

Consider next the case $0<y_0<1$. The stability conditions are given by
\bea
AFC2: & y=0 & Ay_0^{1/2}<8/15 \nn \\
CC2: & 5A(y_c^2+3y_0^2)/4 = (y_0+y_c)^{3/2}+ (y_0-y_c)^{3/2} 
                              & 8/15<Ay_0^{1/2}<2\sqrt{2}/5  \nn \\
CC1: & y_c = 5A(y_0+y_c)^{3/2}/8 & Ay_0^{1/2}>2\sqrt{2}/5  \label{min} \\
FC1: & y = 1 & A(1+y_0)^{3/2}>8/5, \nn
\eea
where $AFC2$, $CC2$, $CC1$ and $FC1$ stand for antiferromagnetic two band 
conducting, canted two band conducting, canted one band conducting and 
ferromagnetic one band conducting respectively.
Notice that $AF$ and canted phases do not compete among them, but only with 
the $F$ phase. The curves providing the boundary between the different phases 
are given by
\bea
AFC2-FC1: & A[(1+y_0)^{5/2} - 2y_0^{5/2}] = 2 & 0<y_0<0.127195 \nn \\
AFC2-CC2: & Ay_0^{1/2}=8/15 & 0.127195<y_0<1 \nn \\
CC2-FC1: & 5A(y_2^2+3y_0^2)/4 = (y_0+y_2)^{3/2} + (y_0-y_2)^{3/2} 
                                    & 0.127195<y_0<0.168457 \nn \\  
CC2-CC1: & Ay_0^{1/2} = 2\sqrt{2}/5 & 0.168457<y_0<1 \\
CC1-FC1: & 5A(y_0+y_1)^{3/2}/8 = y_1 & 0.168457<y_0<0.5 \nn \\
CC1-FC1: & 5A(1+y_0)^{3/2}/8 = 1 & 0.5<y_0<1,  \nn
\eea
where $y_1$ and $y_2$ are given implicitly by the equations
\bea
\lefteqn{ 
[(1+y_0)^{5/2} - (y_0+y_2)^{5/2} - (y_0-y_2)^{5/2}] 
         [(y_0+y_2)^{3/2} + (y_0-y_2)^{3/2}] =
{5 \over 2} (1-y_2^2)(y_2^2+3y_0^2)} \qquad\qquad & & \nn \\
& & \\
\lefteqn{ (y_1+y_0)^{5/2} + 2(1+y_0)^{1/2}(y_1+y_0)^2 + 3(1-y_0)(y_1+y_0)^{3/2} }
\qquad\qquad  \nn \\
& & \mbox{} + 4(1-2y_0)(1+y_0)^{1/2}(y_1+y_0) - 8y_0(1+y_0)(y_1+y_0)^{1/2} 
                                        - 4y_0(1+y_0)^{3/2} = 0. \nn
\eea
The outcome is plotted in fig.~\ref{fig:diagrama}.

\begin{figure}[t]
\begin{center}
\epsfig{file=diagrama.eps,width=12cm,height=10cm,
        bbllx=72,bblly=246,bburx=540,bbury=545}
\end{center}
\caption{
Phase diagram in the $(y_0,A)$ plane. The thick solid line corresponds to first
order transitions whereas the remaining solid lines to second order ones. 
The dotted and dashed dotted lines are the upper stability boundaries for the 
CC1 and CC2 phases respectively.
The two dashed lines are the boundaries for the reliability of our model for 
$z\vert J_H \vert  M / 2(J_{AF}a^3M^2) \sim 50$ and 
$z\vert J_H \vert  M / 2(J_{AF}a^3M^2) \sim 200$ respectively. Only the part 
of the phase diagram to the left of the corresponding dashed line is 
trustworthy in each case. }
\label{fig:diagrama}
\end{figure}

\medskip

Recall that fig.~\ref{fig:diagrama} actually does not plot a phase diagram 
against 
doping but against $y_0$ which is related to the chemical potential rather 
than to the number of conducting fermions or doping. Recall also that 
$V_{{\it eff}}$ is to be regarded as a (zero temperature) grand canonical 
potential rather than as a free energy. The doping is introduced via
\be
x = -a^3\;{\pa V_{\it eff} \over \pa\m} = - {a^3 \over t}\;
                                {\pa V_{\it eff} \over \pa y_0}
\ee
provided that one molecule exists per unit cell with a lattice parameter $a$.
Taking into account (\ref{defs}) the doping corresponding to the different 
phases reads
\bea
AFI: \quad x & = & 0 \nn \\
AFC2: \quad x & = & {z^{3/2} \over 6\pi^2}\; 2 y_0^{3/2} \nn \\
CC2: \quad x & = & {z^{3/2} \over 6\pi^2}\; 
            [(y_0+y_c)^{3/2} + (y_0-y_c)^{3/2}] \label{dop} \\
CC1: \quad x & = & {z^{3/2} \over 6\pi^2}\;(y_0+y_c)^{3/2} \nn \\
FC1: \quad x & = & {z^{3/2} \over 6\pi^2}\;(1+y_0)^{3/2}. \nn
\eea
where the $y_c$ for the $CC2$ and $CC1$ phases are given in (\ref{min}).

\medskip

These expressions for the doping permit us to establish that all our phases
are thermodynamically stable, unlike the ones observed in 
ref.~\cite{Paco2,Kagan}.
This is easily proven from the stability condition $\pa\m / \pa x >0$. For
the $F$ and $AF$ phases this is trivially obtained, whereas canted phases
are stable if they are below the curves
\bea
CC2: & & {5Ay / 3} = (y_0+y)^{1/2} - (y_0-y)^{1/2}  \nn \\
& & \qquad \qquad \qquad y^2 - 5y_0^2 + 4y_0(y_0^2-y^2)^{1/2} = 0  
                                                 \qquad (y<y_0)  \\
CC1: & & Ay_0^{1/2} = {16 / 15\sqrt{3}}. \nn
\eea
This is always the case as it is
shown 
in
fig.~\ref{fig:diagrama}
where we have plotted the two curves.

\medskip

Once we have the expressions (\ref{dop}) for the doping it is straightforward
to translate fig.~\ref{fig:diagrama} to a more conventional phase diagram where 
the doping, $x$, appears in one of the axes. This is given in 
fig.~\ref{fig:fases} (recall $z=6$).

\begin{figure}[t]
\begin{center}
\epsfig{file=fases.eps,width=12cm,height=10cm,
        bbllx=72,bblly=246,bburx=540,bbury=545}
\end{center}
\caption{
Phase diagram in the $(x,A)$ plane. $PSi$ ($i=1,2,3,4$) indicates the new 
regions where the phases at
their boundary may coexist. The $x=0$ axis corresponds to the $AFI$ phase.
The two dashed lines are the boundaries for the reliability of our model for 
$z\vert J_H \vert  M / 2(J_{AF}a^3M^2) \sim 50$ and 
$z\vert J_H \vert  M / 2(J_{AF}a^3M^2) \sim 200$ respectively. Only the part 
of the phase diagram to the left of the corresponding dashed line is 
trustworthy in each case. }
\label{fig:fases}
\end{figure}

\medskip

It is interesting to notice that in fig.~\ref{fig:fases} new regions arise, 
which we have denoted $PSi$ ($i=1,2,3,4$), between the $FC1$ and the $AFI$, 
$AFC2$, $CC2$ and $CC1$ phases respectively. This is due to the fact that 
the thick solid line separating $FC1$ and $AFI$, $AFC2$, $CC2$ and $CC1$ 
in fig.~\ref{fig:diagrama} corresponds to a first order phase transition. 
Along this line two stable inequivalent minima have the same energy and 
the chemical potential cannot be traded by the doping. These regions are 
likely to consist of coexisting domains where the two phases at the boundary 
are realised (phase separation) \cite{Dagotto}. $AFI$ and $FC1$ would coexist 
in $PS1$, as it has been observed in recent works \cite{Paco2,Kagan}. $FC1$ 
and $AFC2$, $CC2$ and $CC1$ would coexist in $PS1$, $PS2$ and $PS3$ 
respectively. These three last possibilities of phase separation have not 
been found before.

\medskip

As mentioned in section~\ref{potential}, the fact that for $t=0$ we do
not permit conductivity restricts the values that the chemical potential takes
to $y_0 < y_0^{max} = \g/2$. By substituting this expression in $A$ we obtain
\be
A = {2z^{1/2} \over 15\pi^2}\;
{z \vert J_H \vert M \over 2(J_{AF}a^3M^2)}\; y_0^{max}.
\ee
which gives the boundary of validity for our results.
It turns out to be a straight line in fig.~\ref{fig:diagrama}
provided that $J_{AF}$ and $J_H$ remains constant as $y_0^{max}$ moves, which 
can be straightforwardly translated to fig.~\ref{fig:fases}. 
Only the phase diagram to the left of this curve is trustworthy.

\medskip

We take for the coupling constants $t/(J_{AF}a^3M^2) \sim 10-20$ and
$z\vert J_H\vert M /2(J_{AF}a^3M^2) \sim 50-200$, which is compatible with 
the values given in the literature. For these values $A \sim 1-2$, and the 
two extreme validity curves are displayed as dashed lines in
fig.~\ref{fig:diagrama} and fig.~\ref{fig:fases}.


\section{Conclusions}
\indent

We have presented a simple model in the continuum which is able to describe 
the rich phase structure of doped manganites for a wide range of these 
materials.

\medskip

We have assumed an underlying cubic crystal for simplicity. 
Nevertheless, the ortho\-rhombic distortion can be easily accommodated by
the following simple changes in the physical parameters: 
$m^3 \rightarrow m_x m_y m_z$, $a^3 \rightarrow abc$,
$J_{AF} \rightarrow J_x + J_y + J_z$ and $t \rightarrow t_x + t_y + t_z$. 
In practice this does not modify our results since it would only lead to a 
different $A$, which is anyway a free parameter in our phase diagrams. This 
fact also suggests that the structural transitions that these materials 
undergo when increasing the doping \cite{Ramirez} 
are not essential in order to understand the $F-AF$ and $I-C$ transitions. 

\medskip

An important feature of our results is that the two canted phases that we 
observe are stable against phase separation, unlike in some previous works 
\cite{Paco2,Kagan}. We also observe regions in the phase diagram where phase 
separations of several kinds may occur. If we plug realistic values for the
physical parameters we find $A \sim 1-2$. Within this range the following 
sequences of phases are possible upon increasing $x$: (i) $AFI-PS1-FC1$, 
(ii) $AFI-AFC2-PS2-FC1$, (iii) $AFI-AFC2-CC2-PS3-FC1$, 
(iv) $AFI-AFC2-CC2-CC1-PS4-FC1$. Recall also that in $PS3$ and $PS4$ 
ferromagnetic and canted phases coexist, which is a situation that has 
not been contemplated in previous works. This may explain some 
controversial results obtained by different authors.

\medskip
 
Let us also mention that the two fermion fields $\psi_1(x)$ and $\psi_2(x)$ 
accommodate the $e_{g}$ doublet in our model. Indeed in the $AF$ phase the two 
lower and two higher eigenvalues (\ref{eigen}) are degenerated. In the $F$ and 
$C$ phases the degeneracy is lifted. This implies that the splitting between 
the two $e_{g}$ levels receives a contribution from the dynamics of the 
conducting fermions in addition to that from the static Jahn-Teller
distortion.

\medskip

The model can be used in the future to study the temperature dependence of 
the phase diagram. Fluctuations due to spin waves in
all the phases (including the canted ones) can also be incorporated \cite{sw}. 
It would also be interesting to see if the model can be generalised to
accommodate the Jahn-Teller distortion dynamically.


\section*{Acknowledgements}
\indent

We are indebted to F. Guinea and L. Brey for introducing us to the physics of 
manganites, and to J. Gonz\'alez for an illuminating discussion in the early 
stages of this work. J.M.R is supported by a Basque Government F.P.I.~grant. 
Financial support from CICYT, contract AEN95-0590 and from CIRIT, contract
GRQ93-1047 is also acknowledged.



\appendix


\Appendix{$\zeta$-function Techniques}
\indent

The $\zeta$-function techniques provide a very efficient way to calculate 
the trace of the logarithm
of operators \cite{Eli}. The $\zeta$-function associated to an
operator $\hat O$ is defined as
\be
\zeta_{\hat O}(s):=tr {\hat O}^{-s}=\sum_{n} \lambda_{n}^{-s}.
\ee
Then
\be
\sum_{n} \log \lambda_{n}= -\left. {d\over ds}\;
                               \zeta_{\hat O}(s)\right\vert_{s=0}.
\ee

\medskip

Consider the operator ${\hat O}$ in (\ref{ohat}). Once the spin 
diagonalisation is performed we only have to consider the 
space-time trace over a generic spin eigenvalue denoted by $\hat O_i$.
Since the real part of the operator $-i{\hat O_i}$ is positive for positive 
energies and negative for negative ones, due to the term $i\epsilon \o$ in
(\ref{eigen}), it is convenient to consider the integral form of 
$\zeta_{\hat O}(s)$  over positive and negative energies separately.
\beasn
tr[({\hat O_i}\theta(-\o))^{-s}] & = &
{(-i)^{-s} \over \G(s)} \int_0^{\infty}{ d\t \t^{s-1}
             \int_{-\infty}^0{{dw \over 2\pi}{d^3{\bf k} \over (2\pi)^3}
\; e^{-i O_i(q)\t}}\; VT}  \\
tr[({\hat O_i}\theta(\o))^{-s}] & = &
{i^{-s} \over \G(s)} \int_0^{\infty}{ d\t \t^{s-1}
             \int_0^{\infty}{{dw \over 2\pi}{d^3{\bf k} \over (2\pi)^3}
\; e^{i O_i(q)\t}}\; VT}.  
\eeasn

\medskip

After the energy and momentum integration we obtain the expressions
\beasn
tr[({\hat O_i}\theta(-\o))^{-s}] & = & 
{VT \over 16\pi} \left({2m \over \pi}\right)^{3/2} {\G(s-5/2) \over \G(s)}
                            \;  (-i)^{-s-5/2} (-i \O_i)^{s+5/2}  \\
tr[({\hat O_i}\theta(\o))^{-s}] & = & 
- {VT \over 16\pi} \left({2m \over \pi}\right)^{3/2} {\G(s-5/2) \over \G(s)}
                            \;  (-i)^{s+5/2} (i \O_i)^{-s+5/2}.
\eeasn
We need the derivative of the above with respect to $s$ at $s=0$. The
presence of $1/\G(s) \sim s$ makes the evaluation very easy, giving rise to
\be
-\left. {d\over ds}\;\zeta_{\hat O}(s)\right\vert_{s=0} =
{VT (2m)^{3/2} \over 30 \pi^2} 
\left[ i^{5/2} (-i \O_i)^{5/2} - (-i)^{5/2} (i\O_i)^{5/2} \right].
\ee

\medskip

The expression between square brackets vanishes when $\O_i>0$, i.e., when
the chemical potential is bellow the energy of the $i$-th state, and is non
zero when $\O_i<0$, i.e., when the chemical potential is above the energy of
the $i$-th state. This leads to the effective potential 
(for $\m <0$, $y_0 < y_0^{max}< 1$)
\bea
V_{{\it eff}} & = & V_0 \left[ (2y^2-1) - {A \over \g^{5/2}}
\left[ \left( \sqrt{1 + {2\g y \over 1+\g^2}} - 
  \sqrt{1 - {2\g y_0 \over 1+\g^2}}\; \right)^{5/2} \theta(y_0+y) 
                                                    \right. \right.\nn \\
& & \mbox{} \phantom{V_0 \left[ (1-2x^2) - A \left[ \right. \right. } + 
      \left. \left. \left( \sqrt{1 - {2\g y \over 1+\g^2}} - 
         \sqrt{1 - {2\g y_0 \over 1+\g^2}}\; \right)^{5/2} \theta(y_0-y) 
                                                      \right] \right],
\eea
where $y=cos(\theta/2)$, whereas $\g$, $y_0$, and $V_0$ are 
defined in (\ref{el}), (\ref{y_0def}) and (\ref{defs}) respectively.
\be
A = {(2m)^{3/2} \over 15\pi^2 J_{AF}M^2} 
               \left(t \;\sqrt{1 + \g^2} \right)^{5/2}.
\ee

\medskip

Eq. (\ref{eigena}) follows from the above by keeping only terms linear in $\g$.


\end{document}